\documentclass[letterpaper, 10 pt, journal, twoside]{support/IEEEtran}
\usepackage{times}
\usepackage[pdftex]{graphicx}
\usepackage{subfigure}
\usepackage{amsmath,amssymb,amsopn,amstext,amsfonts}
\usepackage{cancel}
\usepackage[space]{cite}
\usepackage{soul}

\usepackage{balance}
\usepackage{color}
\usepackage{mathtools}
\usepackage{algorithm}
\usepackage{algorithmic}
\usepackage{bm}

\usepackage{diagbox}
\usepackage{float}
\usepackage{epstopdf}
\usepackage{url}
\usepackage{multirow}
\usepackage{tikz}
\usepackage[linkcolor=black,citecolor=black,urlcolor=black,colorlinks=true]{hyperref}

\soulregister\cite7
\soulregister\citep7
\soulregister\citet7
\soulregister\ref7
\soulregister\it7
\soulregister\pageref7

\graphicspath{{figures/}}
\DeclareGraphicsExtensions{.pdf,.png,.jpg,.eps}
\IEEEoverridecommandlockouts

\title{\LARGE \bf Learning Pugachev's Cobra Maneuver for Tail-sitter UAVs Using Acceleration Model}
\author{Wei Xu$^{1}$, Fu Zhang$ ^{1} $\vspace{-0.0cm}
\thanks{Manuscript received September 9, 2019; revised December 19, 2019; accepted February 10, 2020. This paper was recommended for publication by Editor Jonathan Roberts upon evaluation of the Associate Editor and Reviewers' comment. This work was supported in part by a DJI donation and in part by a HKU seed fund for new staff.}
\thanks{$^{1}$authors are with the Mechatronics and Robotic Systems (MaRS) Laboratory, Department of Mechanical Engineering, University of Hong Kong, China.
		{\{\tt\small xuweii, fuzhang\}@hku.hk}}
\thanks{Digital Object Identifier (DOI): see top of this page.}
\vspace{-0.5cm}}

\begin{document}
	\maketitle
	\begin{abstract}
The Pugachev's cobra maneuver is a dramatic and demanding maneuver requiring the aircraft to fly at extremely high Angle of Attacks (AOA) where stalling occurs. This paper considers this maneuver on tail-sitter UAVs. We present a simple yet very effective feedback-iterative learning position control structure to regulate the altitude error and lateral displacement during the maneuver. Both the feedback controller and the iterative learning controller are based on the aircraft acceleration model, which is directly measurable by the onboard accelerometer. Moreover, the acceleration model leads to an extremely simple dynamic model that does not require any model identification in designing the position controller, greatly simplifying the implementation of the iterative learning control. Real-world outdoor flight experiments on the ``Hong Hu" UAV, an aerobatic yet efficient quadrotor tail-sitter UAV of small-size, are provided to show the effectiveness of the proposed controller.
	\vspace{-0.6cm}\end{abstract}
	\section{Introduction}
\IEEEPARstart{V}{ertical} takeoff and landing (VTOL) UAVs have caught increasing attention these years due to their unique ability to take off and land vertically like a rotorcraft and achieve efficient long-range level flight like a fixed-wing aircraft~\cite{gandhi2019practical}. Several VTOL UAVs, including tilt-rotor\cite{chen2019design}, tilt-wing\cite{rohr2019attitude} and tail-sitter\cite{lyu2017design}, are proposed and researched. Among those configurations, the tail-sitter VTOL has probably the most concise structure that contains no tilting mechanism\cite{mccormick1999aerodynamics}. Moreover, with the optimal design of its aerodynamics and propulsion systems~\cite{gu2018coordinate}, a tail-sitter UAV achieves superior flight efficiency. These features make a tail-sitter UAV ideally suitable for almost all field applications where maneuverability and flight range are two indispensable elements.
	
\begin{figure}[t]
		\vspace{0.0cm}
		\begin{center}
			{\includegraphics[width=1\columnwidth]{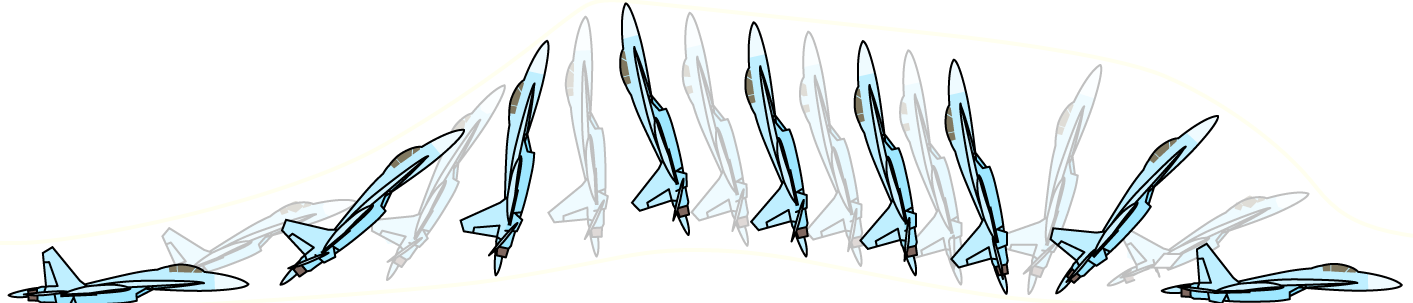}}
		\end{center}
		\vspace{-0.3cm} 
		\caption{\label{fig:cobra-intro}The Pugachev's Cobra Maneuver \protect \footnotemark[2].}
		\vspace{-0.5cm}
	\end{figure}

The Pugachev's Cobra maneuver is a long-existing maneuver in fixed-wing aircraft, where the aircraft flying at high speed suddenly rises its nose momentarily to the vertical orientation and then drops it back to the normal angle, as shown in Fig. \ref{fig:cobra-intro}. Achieving the Pugachev's Cobra maneuvers for small-size tail-sitter UAVs has many practical benefits. For example, in a surveying task, it is usually preferable if the UAV could slow down at a particular location of interests, capturing more data (e.g., images), and then accelerate to the next one. Unlike the Pugachev's Cobra maneuvers for fighter jets, where slight altitude increment is not of concern, or even preferable (Fig. \ref{fig:cobra-intro}), it is usually required for a tail-sitter VTOL UAV of civilian-use to constantly maintain its altitude, to avoid any collision with the environments.
	\footnotetext[2]{https://en.wikipedia.org/wiki/Pugachev}
	
	Although there are a few prior work on the control of fixed-wing UAVs in post-stall maneuvers, such as\cite{levin2016aggressive,sobolic2009nonlinear,matsumoto2010agile}, none of them directly cope with the Pugachev manuever. Another way is to view this maneuver as a backward transition (from level flight to vertical flight) immediately followed by a forward transition (from vertical flight to level flight). For the control of the transition flight, a large number of researches could be found. In~\cite{knoebel2006preliminary}, three transition controllers were investigated and compared on a model Convair XFY-1 Pogo. They are a simple controller based on a vector-thrust model, a feedback linearization controller, and a model reference adaptive controller (MRAC), respectively. All three controllers did not consider the lateral motion. Pucci \emph{et al}.~\cite{pucci2011nonlinear} proposed an equivalent transformation from the original system dynamics to the one that is independent of the aircraft's attitude, enabling the controller to separately compute the thrust and orientation, which will considerably reduce the controller design complexity. In~\cite{naldi2011optimal}, the minimum time and minimum energy optimization problems of VTOL transition were presented and solved numerically. Oosedo \emph{et al}.~\cite{oosedo2017optimal} discussed three transition strategies: standard PID feedback control, minimizing the transition time, minimizing the transition time while regulating the altitude. The latter two strategies are designed based on the aerodynamic model from the wind tunnel test. Ritz \emph{et al}.\cite{ritz2017global} proposed a global controller for a dual-rotor tail-sitter based on the coordinate flight assumption and also did not focus on the accurate altitude control. Smear \emph{et al}.\cite{smeur2018incremental} implemented an incremental nonlinear dynamic inversion controller for the attitude and position control, which has an altitude error of $2\,m$ in outdoor flight test. In these mentioned works, the altitude change of the tail-sitter UAV in a transition is usually not satisfactory. The best of them~\cite{oosedo2017optimal} has an altitude change of around $40\,cm$. Moreover, these methods usually require to know an accurate aerodynamic model of the UAV, and one needs to conduct costly wind tunnel tests. Furthermore, all these methods are designed for a single transition while the performance of the Cobra maneuver involving two consecutive quick transitions could worsen. 
	
	We consider the direct control of Pugachev's Cobra maneuver with tail-sitter UAVs. Our contributions are threefold: (1) A complete position control system consisting of both altitude control and lateral control, which was usually neglected in the prior work~\cite{stone2004control, knoebel2006preliminary, naldi2011optimal, oosedo2017optimal, pucci2011nonlinear}; The key idea is the use of the acceleration model, which hides the underlying UAV dynamics (e.g., motor delay, flexible modes) from the position controller design by using the UAV attitude and body-X acceleration as the virtual control. These virtual control are then tracked by underlying attitude and acceleration controllers in\cite{xu2019full}. (2) An iterative learning controller (ILC) that improves the Cobra maneuver's altitude accuracy through past iterations. Compared to existing ILC methods~\cite{purwin2009performing, schollig2009, mueller2012} and transition controllers of tail-sitters~\cite{oosedo2017optimal,pucci2011nonlinear,smeur2019incremental}, a notable advantage of our method is that it builds on the acceleration model used in the baseline position controller and does not require to identify any of the aircraft model parameters (e.g., aerodynamic coefficients), greatly simplifying its design and implementation on actual systems; (3) Implementation of the whole algorithm on actual tail-sitter UAVs (see Fig. 2) and accomplishment of the lowest altitude change (e.g., less than $30\,cm$, which is within the noise level of typical altitude measurement) reported in the literature so far.
	
The remainder of this paper is organized as follows. Section II will introduce the platform and the acceleration model. The detailed design of the controller will be described in section III. The iterative learning control algorithm will be described in section IV. Experimental results are provided in section IV. Finally, section V draws the conclusion.
	\begin{figure}[t]
		\vspace{0.0cm}
		\begin{center}
			{\includegraphics[width=1.0\columnwidth]{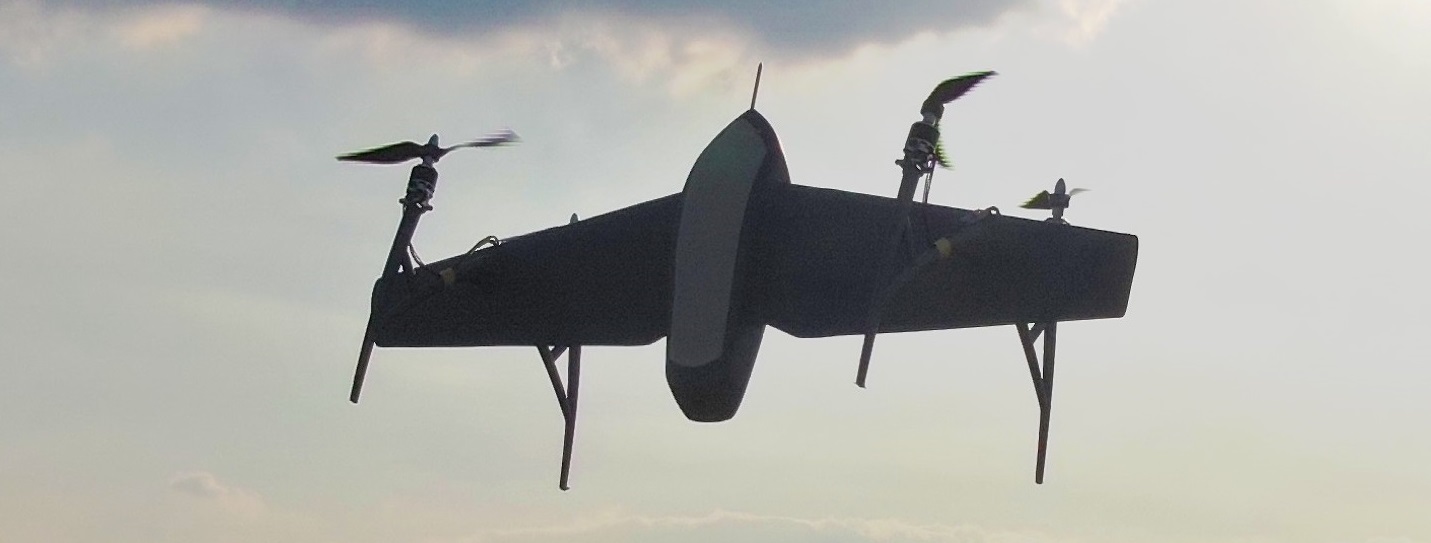}}
		\end{center}
		\vspace{-0.35cm}
		\caption{\label{fig:objective}The Hong Hu quadrotor tail-sitter UAV.}
		\vspace{-0.4cm}
	\end{figure}
	
	\section{System Configuration}
	
	Our tail-sitter UAV is called ``Hong Hu" UAV, which consists of a trapezoidal wing with MH-115 airfoil for improved aerodynamic efficiency, a fuselage containing most of the avionics, and four landing gears with four rotors. Unlike conventional fixed-wing aircraft, the UAV does not have any aileron or elevator, making the propeller differential thrusts the only source of control moments. The lack of control flaps and associated servo motors bring several benefits, such as reduction of dead weight, an increase of flight efficiency due to the removal of gaps between wing and flaps, and simplified controller design without carefully allocating the actuation among rotors and ailerons. The Hong Hu UAV is manufactured with carbon fiber, which leads to improved stiffness and strength of the structure. The flight tests show that the high-speed level flight of the Hong Hu UAV is five times more efficient than that in hovering~\cite{gu2018coordinate}, while still being very portable with a $0.9\,m$ wingspan. The full specifications of ``Hong Hu" UAV  are specified in the Table~\ref{table:config}.
	
	All the flight controller software is running on a Pixhawk 4 Mini controller board\cite{autopilotpixhawk}. The flight sensors include an onboard inertial measurement unit (IMU), a GPS/GNSS module, and a pitot tube airspeed sensor~\cite{lyu2017hierarchical}. All the sensor data are fused within an extended Kalman filter~\cite{meier2012pixhawk} to produce the position, attitude, and velocity estimates.
	\begin{table}[t]
		\vspace{0.15cm}
		\newcommand{\tabincell}[2]{\begin{tabular}{@{}#1@{}}#2\end{tabular}}
		\caption{Aircraft Specification}
		\label{table:config}
		\vspace{-0.15in}
		\begin{center}
			\begin{tabular}{|c|c|}
				\hline
				Battery& $5000\, mAh$\\            
				\hline
				Wingspan& $0.90\,m$\\
				\hline
				Cruise Speed&$19\,m/s$\\
				\hline
				Payload&$0.2\,kg$\\
				\hline
				Weight (with payload)&$1.8\,kg$\\
				\hline
				Range (with payload)&$30\,km$\\
				\hline
			\end{tabular}
		\end{center}
		\vspace{-0.5cm}
	\end{table}
	\subsection{Definition of Coordinate Frames}
\begin{figure}[h]
		\vspace{-0.25cm}
		\begin{center}
			{\includegraphics[width=0.8\columnwidth]{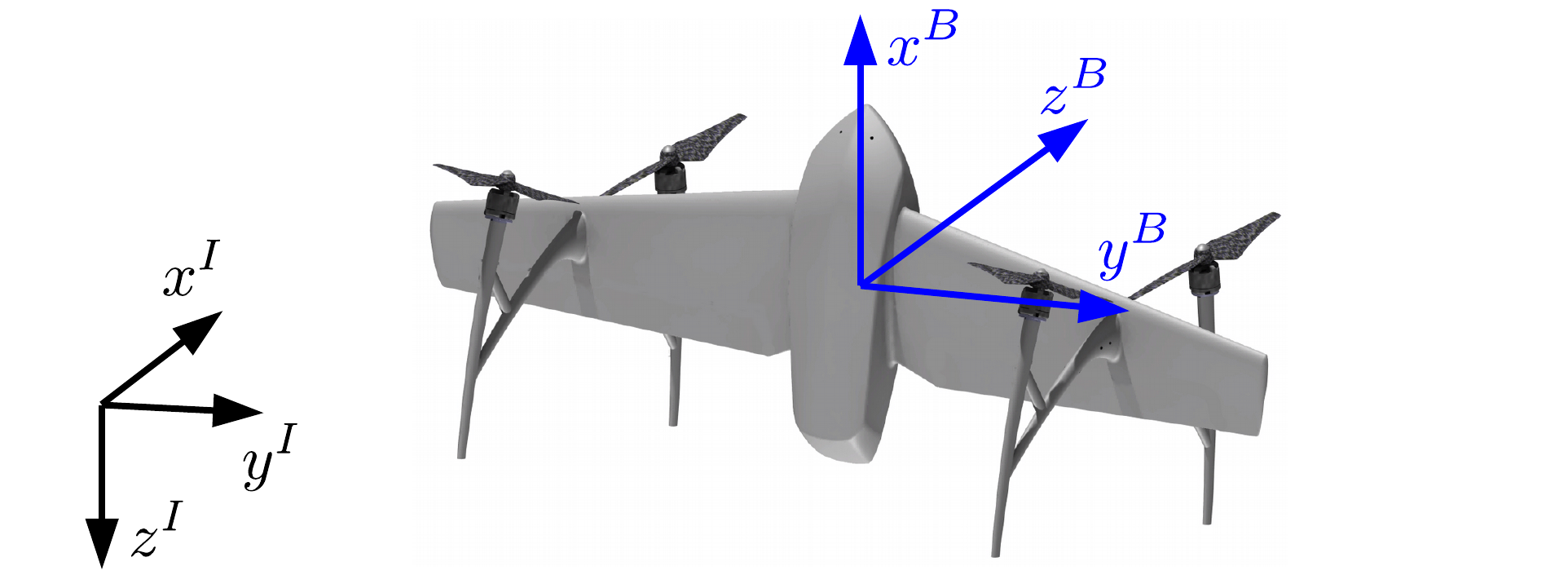}}
		\end{center}
		\vspace{-0.5cm}
		\caption{\label{fig:coordinate}Definations of the body and inertial frames.}
		\vspace{-0.5cm}
	\end{figure}

	The inertial frame ($\bm x^{\mathcal{I}}\bm y^{\mathcal{I}}\bm z^{\mathcal{I}} $)\protect \footnotemark[3] is defined in Fig. \ref{fig:coordinate}, the axis $\bm  x^{\mathcal{I}} $, $\bm y^{\mathcal{I}} $ and $ \bm z^{\mathcal{I}} $ points to the North, East and Down (NED), respectively. For the body frame, the axis $\bm x^{\mathcal{B}} $ points to the aircraft's nose, and $\bm y^{\mathcal{B}} $ points to the right of the body, then $\bm z^{\mathcal{B}} $ is defined by right-hand rule.\footnotetext[3]{Throughout the text the superscript $^\mathcal{I}$ and $^\mathcal{B}$ will be used to denote the inertial and body frame, respectively.}
	
	\subsection{Acceleration Model}
	The position of the aircraft is denoted as $ \bm p^{ \mathcal{\bm I}}=\left[p_{x}^{\mathcal{I}},\,p_{y}^{\mathcal{I}},\,p_{z}^{\mathcal{I}}\right]^{T} $. The angular velocity vector represented in the body frame is $\bm \omega^{\mathcal{B}} $. The total mass of the aircraft is $ m $. Based on the rigid body assumption and the Newton's second law of motion, translational dynamics of the aircraft can be modeled as follows
	\begin{equation}
	\label{e:dynamics}
	\begin{array}{ll}
	\ddot{\bm p}^{\mathcal{I}}=\bm a^{\mathcal{I}}_g+\bm R\bm a^{\mathcal{B}}
	\end{array}
	\end{equation}
	where $ \bm a^{\mathcal{I}}_g = \left[0,\,0,\,g \right]^T $ is the gravity acceleration, the $ \bm a^{\mathcal{B}}=\left[a_{x}^{\mathcal{B}},\,a_{y}^{\mathcal{B}},\,a_{z}^{\mathcal{B}}\right]^{T} $ is the specific acceleration that comprises of the motor thrusts and all the aerodynamic forces. It should be noticed that the specific acceleration is directly measurable by an accelerometer in IMU~\cite{groves2013principles}. The $\bm R $ is the rotation matrix from body frame to inertial frame. 
	
	Without loss of generality, assume that the desired flight path of the Cobra maneuver is along $\bm x^\mathcal{I}$ direction, which is also the initial direction of $\bm z^\mathcal{B}$ of the aircraft (Fig.\ref{fig:coordinate}). Therefore the initial attitude $\bm R_0$ is as follow:
	\begin{equation}\label{equ:rotationmatrix}
	\begin{aligned}
	\bm R_{0}&=\left[ \begin{matrix} 0 & 0 & 1 \\
	0 & 1 & 0 \\
	-1 & 0 & 0 \end{matrix}\right]
	\end{aligned}
	\end{equation}
	
	During the Cobra maneuver, the UAV transits to level flight by rotating along its body-Y axis by angle $\theta$. In this process, the UAV may deviate from the disired path due to disturbances in the environment (e.g., wind). To correct this deviation, the UAV rotates along its body-Z axis ($\bm z^{\mathcal{B}}$ pointing to the aircraft belly) by angle $\eta$, such that the propeller thrust is projected to the lateral direction to compensate the lateral deviation. As a consequence, the UAV rotation matrix is naturally parameterized by two consecutive rotations $\bm R_y(\theta)$, and $\bm R_z(\eta)$ after the initial attitude $\bm R_0$:
	\begin{equation}
	\label{e:dynamic-3}
	\begin{aligned}
	\bm R &= \bm R_{0} \bm R_{y}(\theta)\bm R_{z}(\eta) \\
	&= \left[ \begin{matrix} -\sin(\theta)\cos(\eta)&-\sin(\theta)\sin(\eta)&\cos(\theta) \\ \sin(\eta)&\cos(\eta)&0 \\
	-\cos(\theta)\cos(\eta)& \cos(\theta)\sin(\eta)&-\sin(\theta) \end{matrix}\right]
	\end{aligned}
	\end{equation}
	
	Putting (\ref{e:dynamic-3}) into (\ref{e:dynamics}) leads to:
	\begin{equation}\label{e:a}
	\begin{aligned}
	\ddot p_y^{\mathcal{I}}&=a_x^{\mathcal{B}}\sin(\eta)+a_y^{\mathcal{B}}\cos(\eta)\\
	\ddot p_z^{\mathcal{I}}&=g-a_x^{\mathcal{B}}\cos(\theta)\cos(\eta)+a_y^{\mathcal{B}}\cos(\theta)\sin(\eta)-a_z^{\mathcal{B}}\sin(\theta)
	\end{aligned}
	\end{equation}
	where the dynamics in inertial X direction is omitted (see section III. A). It is clearly seen from (\ref{e:a}) that, the component $a_l = a_x^\mathcal B \sin(\eta) $ solely affects the lateral motion and the component $a_{xp} = a_x^\mathcal B \cos(\eta)$ solely affects the altitude. These two acceleration components will be used as the virtual control action for the lateral and altitude controllers, respectively, as detailed in the next section. The relation between $a_x^\mathcal{B}$, $a_l$, and $a_{xp}$ are illustrated in Fig.\ref{fig:horiz-force}. Using the new notations, the aircraft model is rewritten as: 
	\begin{figure}[t]
		\vspace{0.1cm}
		\begin{center}
			{\includegraphics[width=0.8\columnwidth]{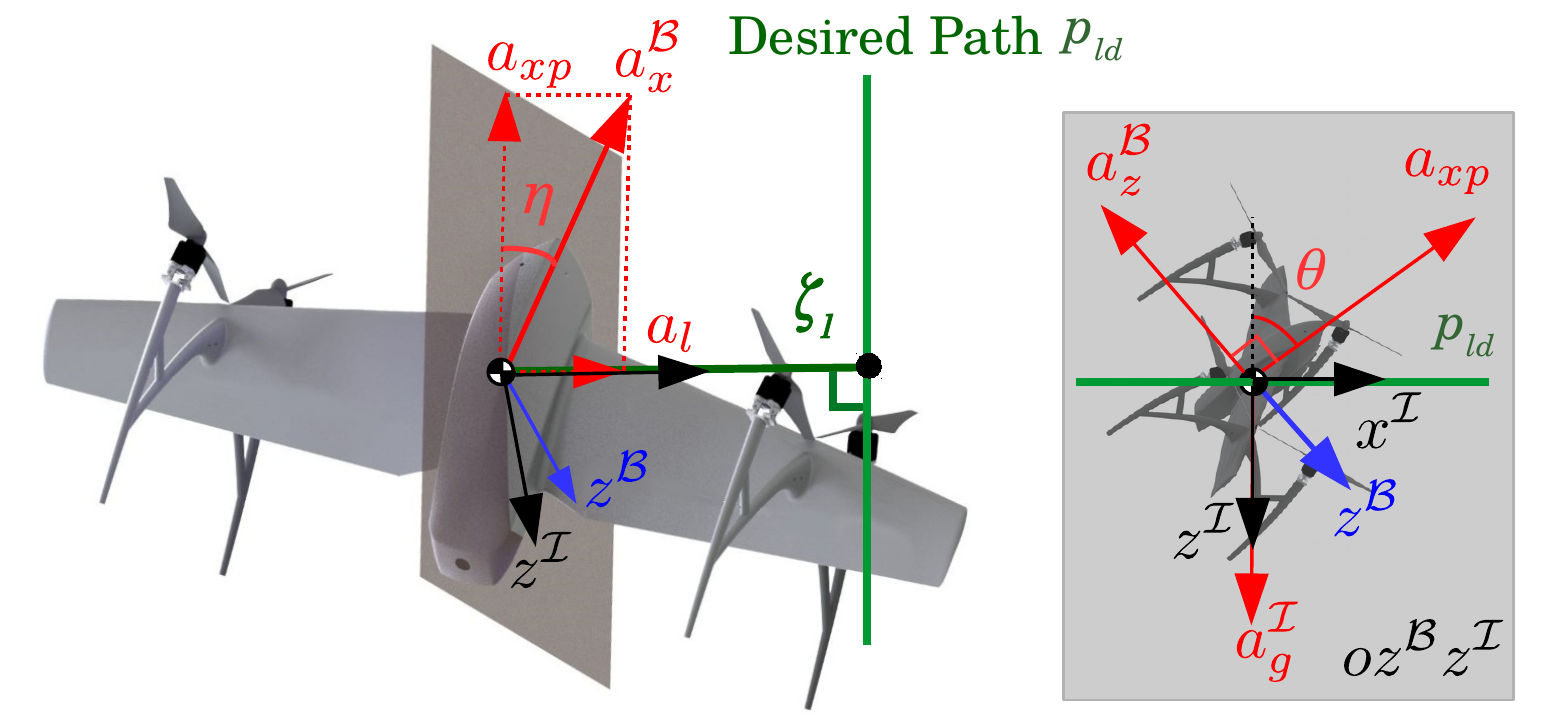}}
		\end{center}
		\vspace{-0.4cm}
		\hspace{2.2cm}(a)\hspace{3.8cm}(b)
		\vspace{-0.2cm}
		\caption{\label{fig:horiz-force}(a) Body-X acceleration $ a_x^{\mathcal{B}} $ and its two orthogonal components: $ a_{xp} $ and $ a_{l} $. It should be noted that $ a_l $ is perpendicular to three vectors: the desired path (as $ a_{l} $ should maximally suppress the lateral deviation), the $ \bm z^{\mathcal{I}} $ axis (as $a_{l}$ is in the horizontal plane), and the $ \bm z^{\mathcal{B}} $ (as $ a_{l} $ is the component of $ a_x^{\mathcal{B}} $ perpendicular to the plane of $\bm o$-$\bm z^{\mathcal{B}}$-$\bm z^{\mathcal{I}} $). $ a_{xp} $ is the component of $ a_x^{\mathcal{B}} $ within the plane $\bm o$-$\bm z^{\mathcal{B}}$-$\bm z^{\mathcal{I}} $. This component will affect the UAV altitude. (b) The accelerations of the UAV in the $\bm o$-$\bm z^{\mathcal{B}}$-$\bm z^{\mathcal{I}} $ plane. $ a_z^{\mathcal{B}}$ is the acceleration in body-Z axis, $a_{xp}$ is the component of $a_x^{\mathcal{B}}$, the acceleration in body-X axis, projected to the $\bm o$-$\bm z^{\mathcal{B}}$-$\bm z^{\mathcal{I}}$ plane, and $\bm a_g^\mathcal{I}$ is the gravity.}
		\vspace{-0.5cm}
	\end{figure}
	\begin{equation}\label{e:accmodel}
	\begin{aligned}
	\ddot p_y^\mathcal{I}&=a_l+a_y^{\mathcal{B}}\cos(\eta)\\
	\ddot p_z^{\mathcal{I}}&=g-a_{xp}\cos(\theta)+a_y^{\mathcal{B}}\cos(\theta)\sin(\eta)-a_z^{\mathcal{B}}\sin(\theta)
	\end{aligned}
	\end{equation}
	
	Equation (\ref{e:accmodel}) is the acceleration model used for the lateral controller and altitude controller design. There are several benefits of this acceleration model: 1) With the acceleration model, the UAV translation dynamics are linear to the two virtual control action $a_{xp}$ and $a_l$, and does not depend on any model parameters (e.g., aerodynamic coefficients, etc.). This greatly simplifies the controller design; 2) The acceleration is easily measurable by IMU sensors, which are almost always available for any UAVs; 3) The body-X acceleration is independently controlled by the collective thrust of the four rotors.
	
	\section{Feedback Controller Design}
	\subsection{Controller Structure}
\begin{figure}[h]
		\vspace{0.0cm}
		\begin{center}
			{\includegraphics[width=1\columnwidth]{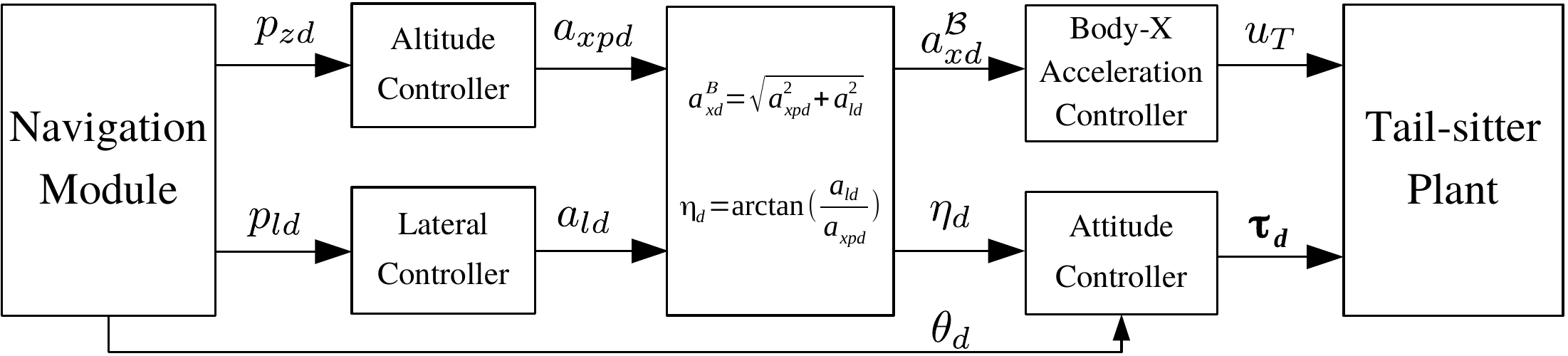}}
		\end{center}
		\vspace{-0.25cm}
		\caption{\label{fig:ctrl_struc}{The proposed controller structure.}}
		\vspace{-0.0cm}
	\end{figure}

The controller structure for the Pugachev’s Cobra maneuver is shown in Fig. \ref{fig:ctrl_struc}. The navigation module generates the desired altitude $ p_{zd} $, which is a constant value for a Pugachev’s Cobra maneuver, and the desired flight path $ p_{ld} $, which is a straight line in the horizontal plane (see Fig. \ref{fig:horiz-force}). It also directly sends the desired pitch angle $ \theta_{d} $ according to a prescribed trajectory. Linearly decreasing/increasing pitch profiles are used to test the designed controllers albeit smoother trajectories could lead to better tracking peformance. In cases where precise forward position (i.e., range) needs to be actively controlled, an additional feedback controller viewing the pitch angle as virtual control action could be employed, which is not considered in this work though.
	
	The altitude controller (section III. B) and lateral controller (section III. C) will respectively use the $a_{xp}$ and $a_{l}$ as their virtual control actions (see Equation (\ref{e:accmodel})) and compute their desired values, $a_{xpd}$ and $a_{ld}$, the detail of each controller will follow in next subsections. With the desired value of the two orthogonal components, $a_{xpd}$ and $a_{ld}$, the total specific acceleration along body-X axis, $a_x^\mathcal{B}$, and the angle $\eta$ are expected to be:
	\begin{equation}
	\label{e:sm_ctrl_3}
	\begin{array}{ll}
	\displaystyle a_{xd}^{\mathcal{B}} = \sqrt{a_{xpd}^2+a_{ld}^2}
	\end{array}
	\end{equation}
	\begin{equation}
	\label{e:sm_ctrl_4}
	\begin{array}{ll}
	\displaystyle \eta_{d} = \arctan\left(\frac{a_{ld}}{a_{xpd}}\right)
	\end{array}
	\end{equation}
	
	The desired body-X acceleration, $a_{xd}^\mathcal{B}$, is then fed to a lower level acceleration controller (section III. D) which computes the collective throttles of the four rotors. The desired rotation angle $\eta_d$, together with the desired pitch angle $\theta_d$, are used to compute the desired attitude: $\bm R_d = \bm R_0\bm R_y(\theta_d)\bm R_z(\eta_d)$, which is then sent to a lower-level attitude controller (section III.D) for attitude tracking. 
	\subsection{Altitude Controller}
	Based on the altitude dynamics in (\ref{e:accmodel}), the proposed altitude controller consists of a feedforward controller compensating any exogenous input and a feedback controller stabilizing the altitude error (see Fig. \ref{fig:vert-ctrl-struc}). The total control action for the altitude controller is:
	\begin{figure}[h]
		\vspace{0.0cm}
		\begin{center}
			{\includegraphics[width=1.0\columnwidth]{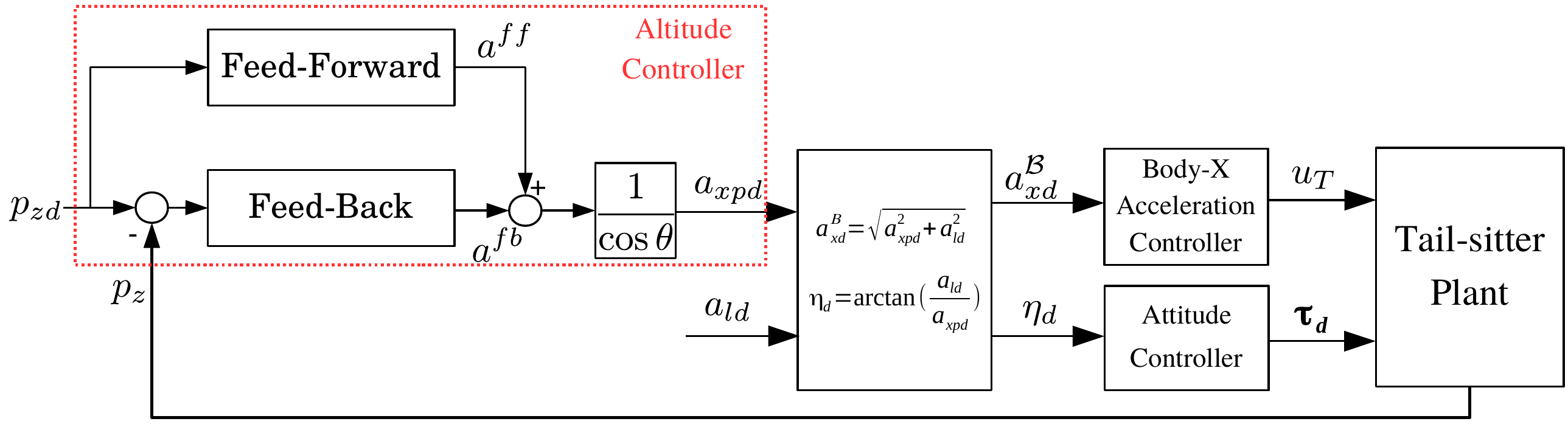}}
		\end{center}
		\vspace{-0.25cm}
		\caption{\label{fig:vert-ctrl-struc}The structure of altitude controller.}
		\vspace{-0.25cm}
	\end{figure}
	\begin{equation}
	\label{e:alt_ctrl_0}
	\begin{array}{ll}
	\displaystyle a_{xpd} = \frac{a^{ff} + a^{fb}}{\cos(\theta)} 
	\end{array}
	\end{equation}
	where,
	\begin{equation}
	\label{e:alt_ctrl_1}
	\begin{array}{ll}
	\displaystyle a^{ff} = -\ddot{p}_{zd}^{\mathcal{I}} + g - a_{z}^{\mathcal{B}}\sin(\theta) + a_y^{\mathcal{B}}\cos(\theta)\sin(\eta)
	\end{array}
	\end{equation}
	\begin{equation}
	\label{e:alt_ctrl_2}
	\begin{array}{ll}
	\displaystyle a^{fb} = k_{p} \xi_p + k_{v}\xi_v
	\end{array}
	\end{equation}
	where the $ p_{zd}^{\mathcal{I}} $ is the desired altitude; $ \xi_p = p_{zd}^{\mathcal{I}} - p_{z}^{\mathcal{I}} $ and $\xi_v = \dot{p}_{zd}^{\mathcal{I}} - v_{z}^{\mathcal{I}} $ is the altitude error and vertical velocity error, respectively; $ k_{p} $ and $ k_{v} $ are the proportional and derivative gains of the feedback controller.
	
	We analyze the stability of the designed controller in the frequency domain\cite{Camacho1997Frequency}. Assume the transient response of the acceleration controller is denoted as a transfer function $G(s)$, then:
	\begin{equation}
	\label{e:xpd_xp}
	\begin{array}{ll}
	\displaystyle a_{xp}(s) = G(s)a_{xpd}(s)
	\end{array}
	\end{equation}
	where $a_{xp}(s)$ and $a_{xpd}(s)$ is the Laplace transform of $a_{xp}$ and $a_{xpd}$. Putting (\ref{e:xpd_xp}) back into the acceleration model (\ref{e:accmodel}) yields:
	\begin{equation}\label{e:vert_close}
	\begin{aligned}
	s^2\xi_{p}(s)=(1-G(s))a^{ff}-G(s)C(s)\xi_p(s)
	\end{aligned}
	\end{equation}
	where $C(s)=k_p+k_v s$ is the transfer function of the feedback controller in (\ref{e:alt_ctrl_2}); The loop transfer function of the system is therefore as follow:
	\begin{equation}\label{e:vert_close}
	\begin{aligned}
	H(s)=\frac{C(s)}{s^2}G(s)
	\end{aligned}
	\end{equation}
	
\begin{figure}[h]
		\vspace{-0.25cm}
		\begin{center}
			{\includegraphics[width=1.0\columnwidth]{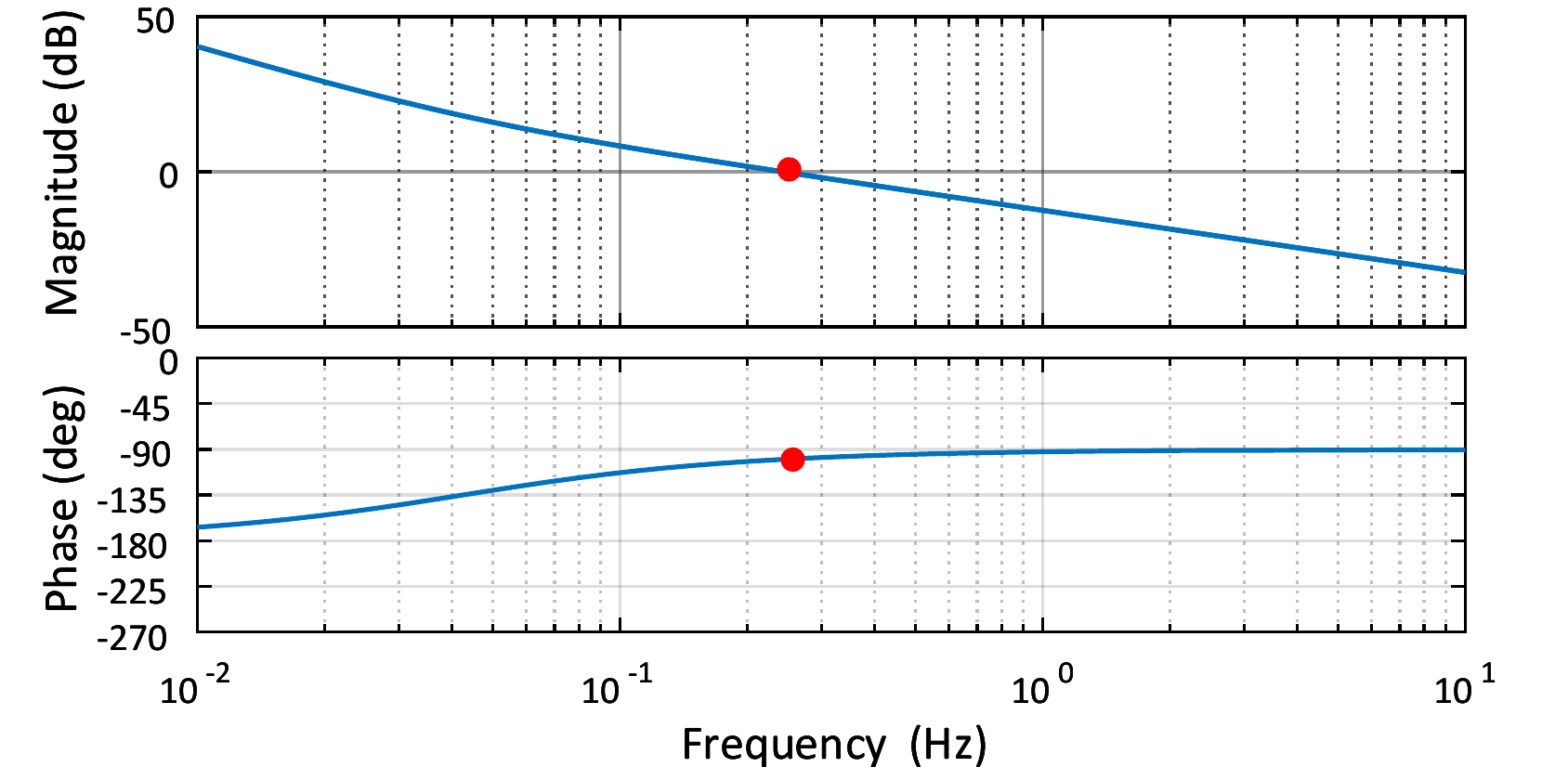}}
		\end{center}
		\vspace{-0.5cm}
		\caption{\label{fig:bode_alt}Bode plot of the nominal loop transfer function of the altitude control loop}
		\vspace{-0.25cm}
	\end{figure}

	To analyze the stability of this system, we investigate the nominal loop transfer function $H_n(s) = C(s)/s^2$. As shown in the Bode plot of $H_n(s)$ in Fig. \ref{fig:bode_alt}, the designed controller has $0.24\,Hz$ bandwidth and a phase margin of $81\,^\circ$, meaning that the position control loop is stable even when adding an up to $0.9\,s$ pure delay into the $G(s)$. This pure delay margin is much higher than needed, showing the sufficient robustness margin of our position controller. Readers may refer to~\cite{Camacho1997Frequency} for more details on the frequency-domain based stability analysis.
	
	\subsection{Lateral Controller}
	Based on the lateral dynamics in (\ref{e:accmodel}), we propose the following proportional controller:
	\begin{equation}
	\label{e:sm_ctrl_1}
	\begin{array}{ll}
	\displaystyle a_{ld} = k_{l}\xi_{l}
	\end{array}
	\end{equation}
	where $\xi_{l}$ is the lateral error, $k_l$ is the gain of lateral controller. This proportional controller acting on the UAV lateral dynamics with the inherent aerodynamic damping effect\cite{katz1988unsteady} ($\displaystyle a_y^\mathcal{B} = k_v^a \dot \xi_l; k_v^a>0 $) forms a second order system, whose stability in the presence of transient response of the lower-level acceleration controllers can be shown following the same procedures as that in the altitude control loop.
	
	\subsection{The Attitude and Body-X Acceleration Controllers}
	The desired attitude $\bm R_d$ and acceleration $a_{xd}^\mathcal B$ computed by the position controllers are tracked by two lower-level controllers, respectively. To achieve this task, we directly use the method in\cite{xu2019full}. The attitude controller is a dual-loop control structure where the outer loop is a quaternion-based attitude controller that operates in the full Special Orthogonal group $SO(3)$. The inner loop consists of three independent angular rate controllers that are designed with loop-shaping techniques. The detailed structure and stability analysis of the attitude controller can be found in\cite{xu2019full}. 
	
\begin{figure}[h]
		\begin{center}
			{\includegraphics[width=0.7\columnwidth]{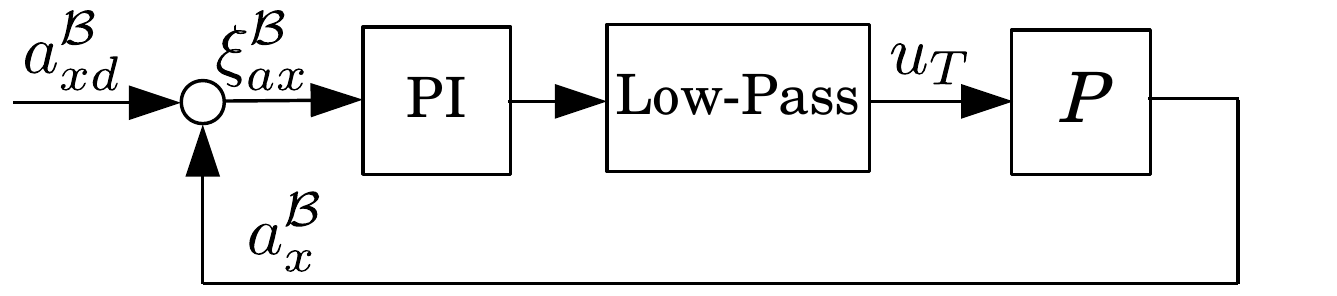}}
		\end{center}
		\vspace{-0.35cm}
		\caption{\label{fig:acc_ctrl}The structure of body-X axis acceleration controller.}
		\vspace{-0.0cm}
	\end{figure}

	The body-X acceleration controller is also detailed in~\cite{xu2019full}. Its main structure is shown in Fig.\ref{fig:acc_ctrl}, where the low pass filter is used to suppress the noise that is prevalent in IMU measurements. The PI controller is used to make sure the command tracking performance. The $P$ denotes the plant dynamic from thrust command $u_T$ to actual body-X acceleration.
	
	It should be noted that although the attitude controller and body-X acceleration controller involves much model identification in the frequency domain, as shown in\cite{xu2019full}, they do not affect the position controller (both the baseline controller in section. III and the ILC controller in section. IV).
	
	\section{Iterative Learning Strategy}
\begin{figure}[h]
		\vspace{0.0cm}
		\begin{center}
			{\includegraphics[width=1\columnwidth]{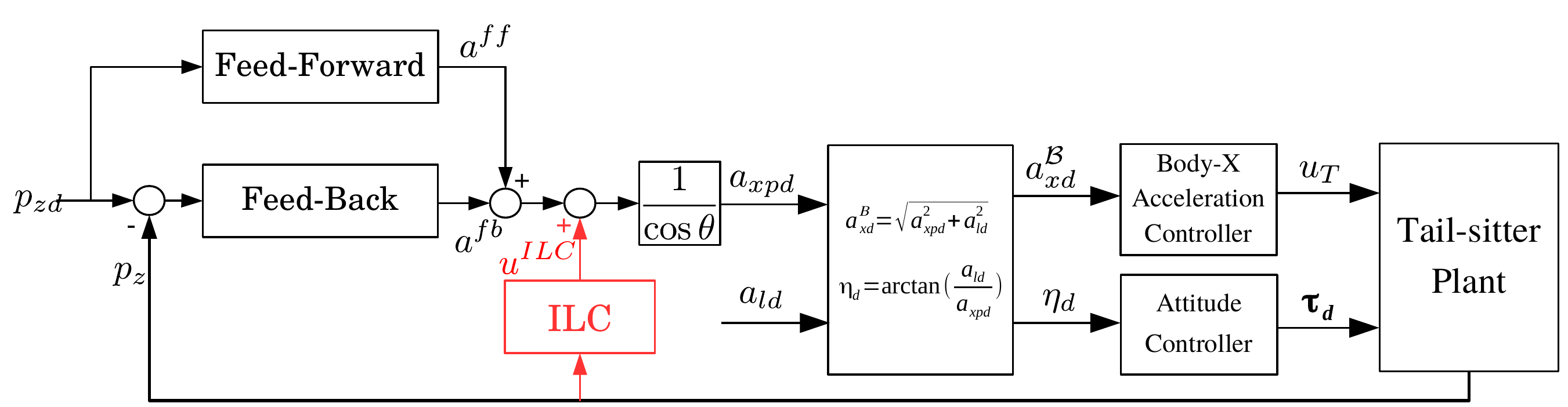}}
		\end{center}
		\vspace{-0.25cm}
		\caption{\label{fig:vert-ctrl-struc-ilc}Adding the ILC correction to the altitude controller.}
		\vspace{-0.25cm}
	\end{figure}

	The attitude controller designed in the previous section, even with the feedforward corrections, usually cannot track the constant altitude command with satisfactory performance due to environmental disturbances and the transient response of the lower level acceleration controller (i.e., $G(s)$). To track the desired path precisely, an iterative learning feedforward controller is further utilized to the altitude controller, as shown in Fig.\ref{fig:vert-ctrl-struc-ilc}.
	\vspace{-0.25cm}
	\subsection{Lifted Domain Model}
	The first step of using the iterative learning algorithm is obtaining a lifted domain model near the desired trajectory. In this paper, the altitude error $ \xi_p $ and the vertical velocity $\xi_v $ in the inertial frame are chosen to compose the state vector $ \bm x $:
	\begin{equation}
	\label{e:state_vector_x}
	\begin{array}{ll}
	\bm x= \left[\xi_p, \xi_v\right]^T,\quad \bm y= \xi_p
	\end{array}
	\end{equation}
	
	When the iterative learning correction $ u^{ILC} $ is added to the altitude controller, (\ref{e:alt_ctrl_0}) is rewritten as:
	\begin{equation}
	\label{e:alt_ctrl_11}
	\begin{array}{ll}
	\displaystyle a_{xpd} = \frac{a^{ff} +a^{fd} + u^{ILC}}{\cos(\theta)}
	\end{array}
	\end{equation}
	
	As shown in\cite{schollig2009}, an ILC uses a nominal model of the closed-loop system to estimate the external disturbances. The error due to the model mismatch is lumpped into disturbances and will be attenuated via iterations. In our design, the nominal model is obtained by neglecting the transient response (i.e., $G(s) = 1$) in (\ref{e:vert_close}). As a result,
	\begin{equation}
	\label{e:linear-1}
	\begin{aligned}
	\displaystyle \dot{\bm x} &= \bm A\bm x + 
	\bm B{u^{ILC}}\\
	\bm y&=\bm C\bm x\\
	\bm A=\left[\begin{matrix}
	0&  1\\ 
	-k_{p}&  -k_{v}\\  
	\end{matrix}\right],&\quad \bm B=\left[\begin{matrix}
	0\\ 
	1\end{matrix}\right],\quad \bm C=[1,0]
	\end{aligned}
	\end{equation}
	
	The corresponding closed-loop state equation in discrete time domain is:
	\begin{equation}
	\label{e:discrete-1}
	\begin{array}{ll}
	\displaystyle \bm x(k+1) &= \bm A_{d}\bm x(k) + \bm B_{d} u^{ILC}(k)\\
	\displaystyle \bm y(k+1) &=\bm C_{d}\bm x(k)
	\end{array}
	\end{equation}
	where $ k\in \left \{ 0,1,...,N \right \} $ is the discrete-time index, $ \Delta t $ is the sampling time, $ \bm A_{d}=\bm I + \bm A\Delta t $, $ \bm B_{d}=\bm B\Delta t $ and $ \bm C_{d} = \bm C $.
	
	In the actual flight, the lifted domain\cite{mueller2012} input vector is denoted as $ \overline{\bm U} = \left[u^{ILC}(0),...,u^{ILC}(N-1)\right]^{T}\in U \subset \mathbb{R}^{N} $, the state vector is denoted as $\overline{\bm X} = \left[\bm x(0)^T,...,\bm x(N-1)^T\right]^{T}\in U \subset \mathbb{R}^{2N}$, the output vector is denoted as $ \overline{\bm Y} = \left[\bm y(0),...,\bm y(N-1)\right]^{T}\in U \subset \mathbb{R}^{N}$. The desired trajectory is denoted as the desired trajectory $\left(\overline{\bm U}_{des}, \overline{\bm X}_{des},\overline{\bm Y}_{des}\right) $. For the ideal Pugachev's Cobra, $ \overline{\bm Y}_{des}=\bm 0_{N\times 1} $. The initial trajectory where no iterative control input is used (i.e. $u^{ILC}(k) = 0, k = 1, 2, ..., N$) is denoted as $\left( \overline{\bm U}_{init} = \bm 0_{N\times 1},\overline{\bm X}_{init},\overline{\bm Y}_{init}\right) $. These lifted domain notations allows us capture the dynamic relation between input, state, and output by a simple linear mapping:
	\begin{equation}
	\label{e:F-mat-1}
	\begin{aligned}
	\displaystyle \overline {\bm Y} &= \bm F\overline {\bm U} + \bm d
	\end{aligned}
	\end{equation}
	where the $\bm F$ is computed by (\ref{e:F-mat-2}), $ \bm d $ is the unknown repetitive disturbance along the trajectory, which is primarily caused by unmodeled dynamics (e.g., transient response of the lower-level acceleration controller) and steady environmental disturbances (e.g., wind, etc.).
	\begin{equation}
	\vspace{-0.15cm}
	\label{e:F-mat-2}
	\begin{aligned}
	\bm F = \left[\begin{matrix}\bm 0 & ... & \bm 0 & \bm 0\\  \bm C_d \bm A_d \bm B_d & ... & \bm 0 & \bm 0\\ \vdots & \vdots & \vdots & \vdots \\ \bm C_d \bm A_d^{N-1} \bm B_d & ... & \bm C_d \bm A_d \bm B_d & \bm 0\\
	\end{matrix}\right]
	\end{aligned}
	\end{equation}
	
	It should be noticed that the model in (\ref{e:F-mat-2}) is exactly known, involving no aerodynamic parameters that require additional effort (e.g., costly wind tunnel test, model identification by flight experiments). This will considerably simplify the following iterative learning controller design, which is a major advantage of our method using the acceleration model.
	
	\subsection{Disturbance Estimation and Input Update}
	An iteration domain Kalman filter is applied to calculate the estimate $ \widehat{\bm d}_{i+1} $ of the disturbance vector $ \bm d $ in (\ref{e:F-mat-1}) after each iteration as (\ref{e:KF-1}), $ i $ indicates the $ i $-th execution of the Pugachev's Cobra maneuver.
	\begin{equation}
	\label{e:KF-1}
	\begin{aligned}
	\displaystyle \widehat{\bm d}_{i+1}=\widehat{\bm d}_{i}+\bm K_{i}\left(\overline{\bm Y}_{i}-\bm F\overline {\bm U}_{i}-\widehat{\bm d}_{i}\right)
	\end{aligned}
	\end{equation}
	where the $ \bm K_{i} $ is the Kalman gain of which the details can be found in\cite{schollig2009}.
	
	Then based on the estimated disturbance (\ref{e:KF-1}), the iterative learning control input to be applied at the next iteration $ \widehat{\bm U}_{i+1} $ should minimize the altitude tracking error predicted by the model (\ref{e:F-mat-1}), i.e., $\bm U_{i+1}$ is solved from the following optimization problem.
	\begin{equation}
	\label{e:update-2}
	\begin{aligned}
	\displaystyle \min_{\overline {\bm U}_{i+1}} &\quad \left \| \bm F\overline {\bm U}_{i+1}+\widehat{\bm d}_{i+1} -\overline{\bm Y}_{des} \right \|_{2} + \alpha  \left \| \overline {\bm U}_{i+1} \right \|_{2}\\
	s.t. & \quad \overline {\bm U}_{i+1} \preceq \bm c_{max}
	\end{aligned}
	\end{equation}
	where the first term stands for the 2-norm of the predicted tracking error, and the second one is the penalty term to the iterative learning control input with the weight $ \alpha $; The lifted vector $ \bm c_{max} $ denotes the maximally allowed $ \overline {\bm U}_{i+1} $.
	
	With the known constant matrix $\bm F$ and disturbance estimate vector $\widehat{d}_{i+1}$, the problem described by (\ref{e:update-2}) is a convex optimization problem\cite{bubeck2015convex}, which can be solved by software tool such as the CVX toolbox of Python.
	\section{Experiments Verification}
	This section presents the experimental results of the proposed controller design and learning scheme applied to the Hong Hu tail-sitter UAV described in section II.
	\subsection{Body-X Acceleration Controller}
\begin{figure}[t]
		\vspace{0.15cm}
		\begin{center}
			{\includegraphics[width=0.95\columnwidth]{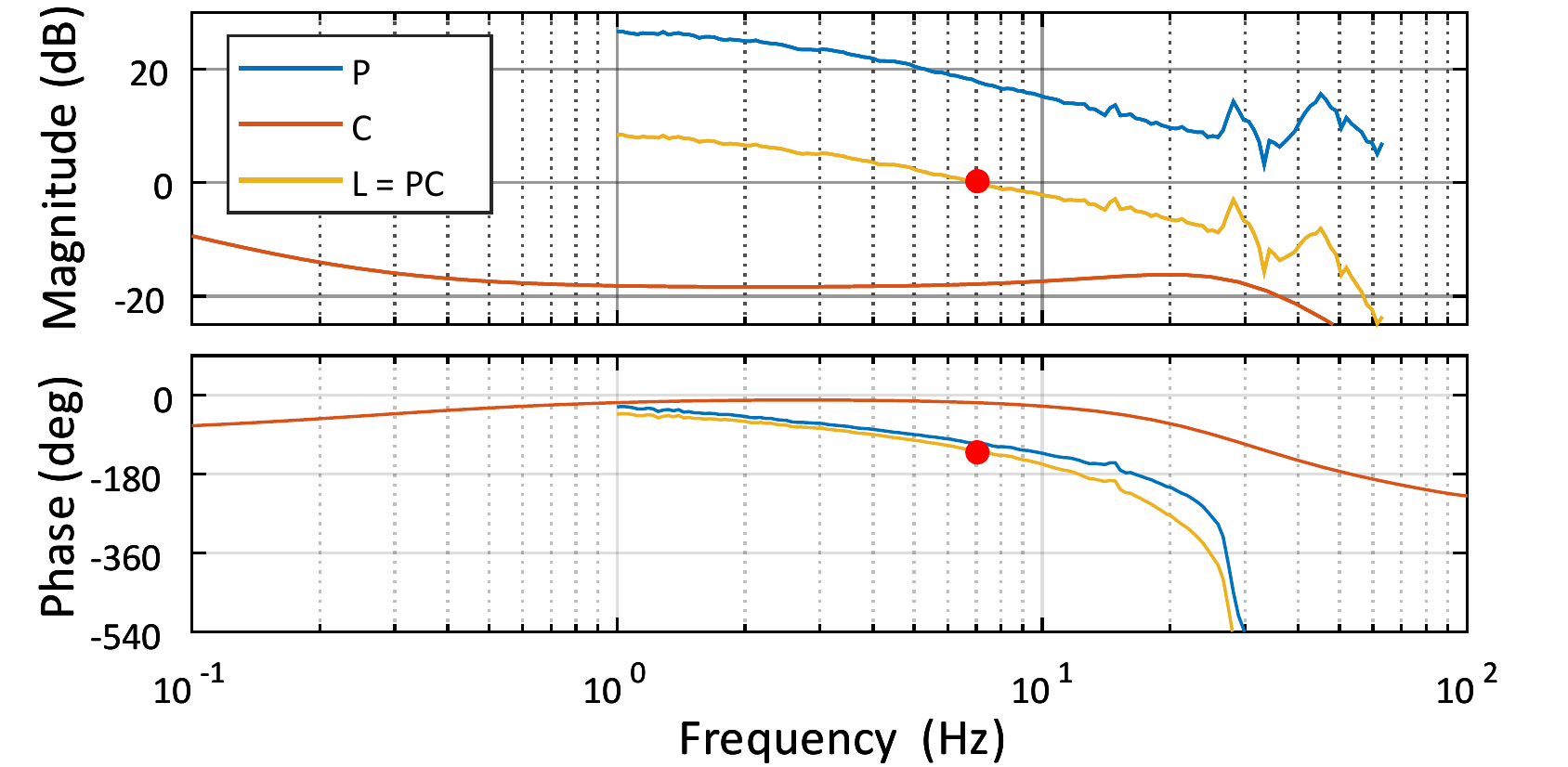}}
		\end{center}
		\vspace{-0.5cm}
		\caption{\label{fig:ACC_LOOP}Plant transfer function, controller transfer function, and the loop transfer function of the acceleration loop}
		\vspace{-0.25cm}
	\end{figure}

	To examine the stability and robustness of the body-X acceleration controller, Fig.\ref{fig:ACC_LOOP} shows the loop transfer function $L=PC$ in the frequency domain, where $ P $ is the actual plant identified from frequency sweeping experiment\cite{xu2019full}, the $ C $ is our designed PI controller with a low-pass filter. The PI controller is used to increase the low-frequency gain of $L$, which compensates for any low-frequency disturbance from the environment or the aircraft itself. The low-pass filter is used to decrease the influence of high-frequency noise, which mainly comes from the propeller disturbance and IMU measurement noise. The bias of the accelerometer is estimated by an EKF with GPS measurements\cite{meier2012pixhawk}.
	
	Based on the loop-shaping technique, which is discussed in detail in our prior work\cite{xu2019full}, the corner frequency of PI controller and low-pass filter is set to $0.9\,Hz$ and $14\,Hz$, respectively. As shown by the loop transfer function $L(s)$ in Fig. \ref{fig:ACC_LOOP}, the acceleration control loop is stable with a phase margin up to $54\,^\circ$, which is in the desired range for typical aircraft systems to ensure sufficient robustness margin. Furthermore, the bandwidth of the acceleration control loop is as high as $7.2\,Hz$, leading to a phase delay of $G(s) = L(s)/(1+L(s))$ nearly zero at $0.24\,Hz$, the crossover frequency of the  loop transfer function of the outer-loop position control (see Fig. \ref{fig:bode_alt}). This nearly zero phase delay is well below the $81\,^\circ$ phase margin of the outer-loop position control and hence guarantees the stability of the designed position controller in section III. B.
	
	\subsection{Lateral Controller}
\begin{figure}[h]
		\vspace{-0.25cm}
		\begin{center}
			{\includegraphics[width=1\columnwidth]{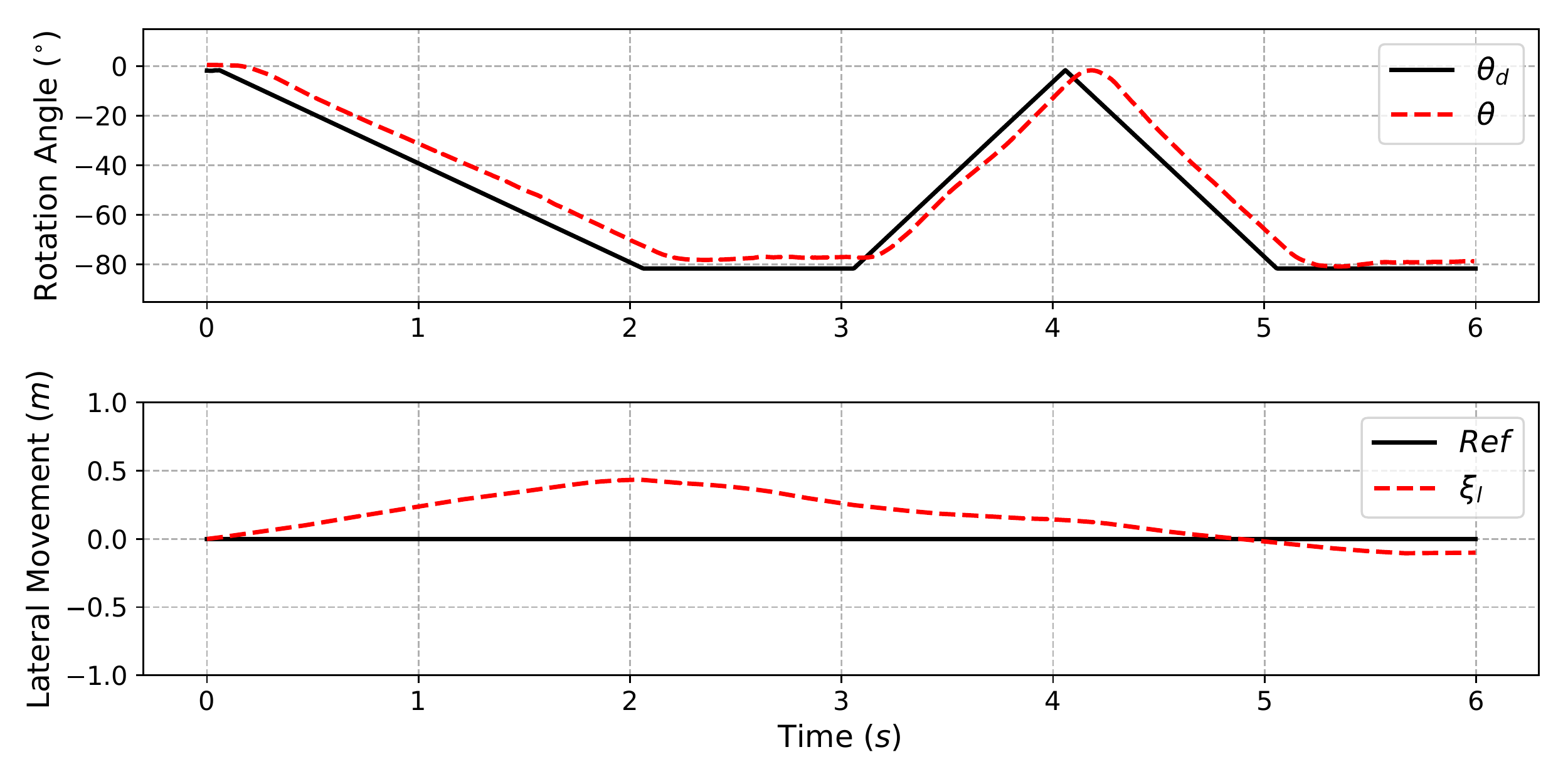}}
		\end{center}
		\vspace{-0.5cm}
		\caption{\label{fig:ATTI_TEST} The attitude response and lateral movement.}
		\vspace{-0.5cm}
	\end{figure}

	For the controller validation experiments, the desired pitch angle in Pugachev's Cobra maneuver is shown in Fig. \ref{fig:ATTI_TEST}. The aircraft will head up to $0\,^\circ$ after the first forward transition. Fig. \ref{fig:ATTI_TEST} also shows the attitude response and lateral movement. It is apparent that the actual $ \theta $ tracks the prescribed trajectory $ \theta_{d} $ very well and does not performs the . Moreover, with the designed lateral controller, the deviation from the flight path is below 1 m. 
	\subsection{Altitude Controller}
\begin{figure}[t]
		\begin{center}
			{\includegraphics[width=0.95\columnwidth]{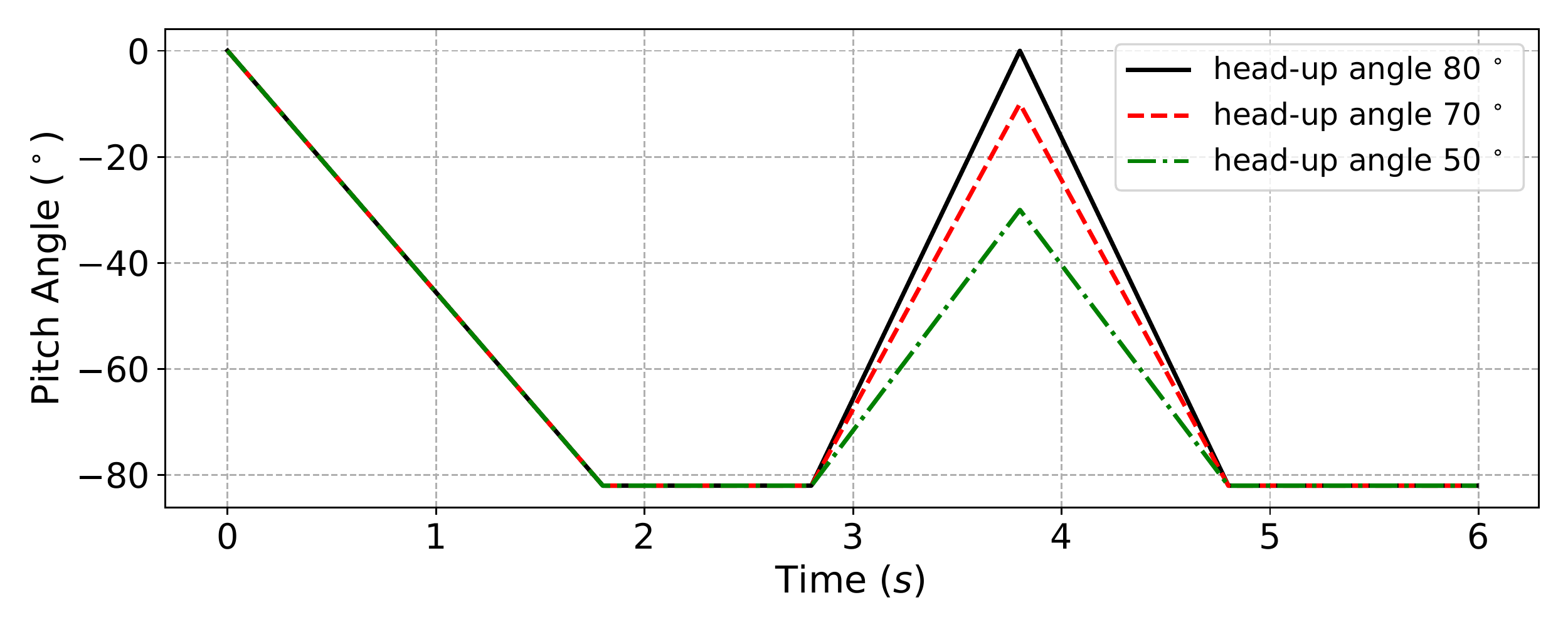}}
		\end{center}
		\vspace{-0.5cm}
		\caption{\label{fig:head_ang} Three different profiles of desired pitch angle implemented in the experiments}
		\vspace{-0.5cm}
	\end{figure}

	We test our control methods on three Cobra maneuvers of different pitch profiles. As shown in Fig. \ref{fig:head_ang}, the three pitch profiles differ in the head-up angle (i.e.,  $80\,^\circ$, $70\,^\circ$, and $50\,^\circ$), all in the post-stalling region.
	
\begin{figure}[h]
		\vspace{-0.cm}
		\begin{center}
			{\includegraphics[width=1.0\columnwidth]{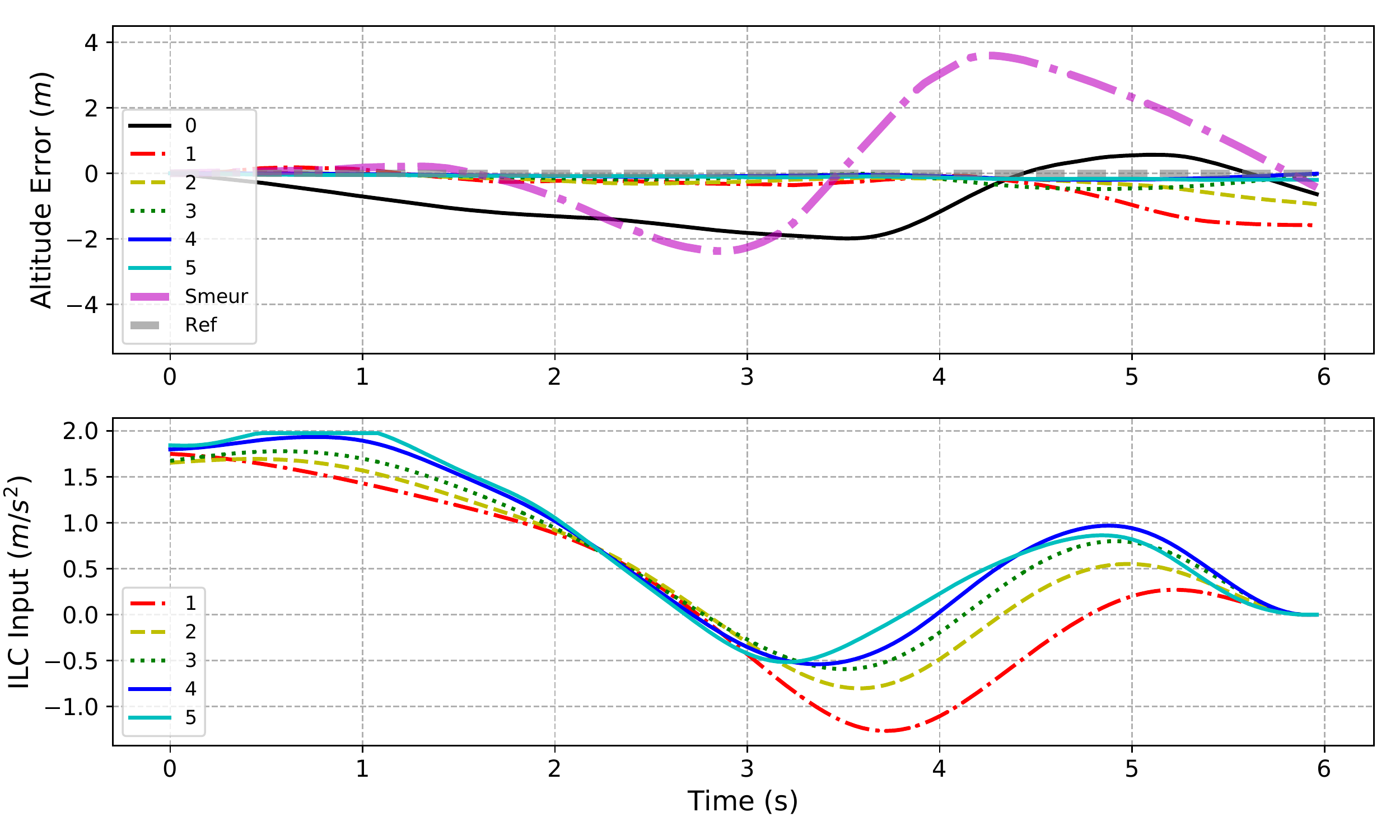}}
		\end{center}
		\vspace{-0.5cm}
		\caption{\label{fig:ILC-test} Altitude error and ILC input during each iteration when the head-up angle is $80\,^\circ$. The index 0 stands for the initial trajectory; ``Ref" stands for the constant altitude setpoint; ``Smeur" stands for the experiment using the method from \cite{smeur2019incremental}.}
		\vspace{-0.25cm}
	\end{figure}

\begin{figure}[h]
		\vspace{0.cm}
		\begin{center}
			{\includegraphics[width=1.0\columnwidth]{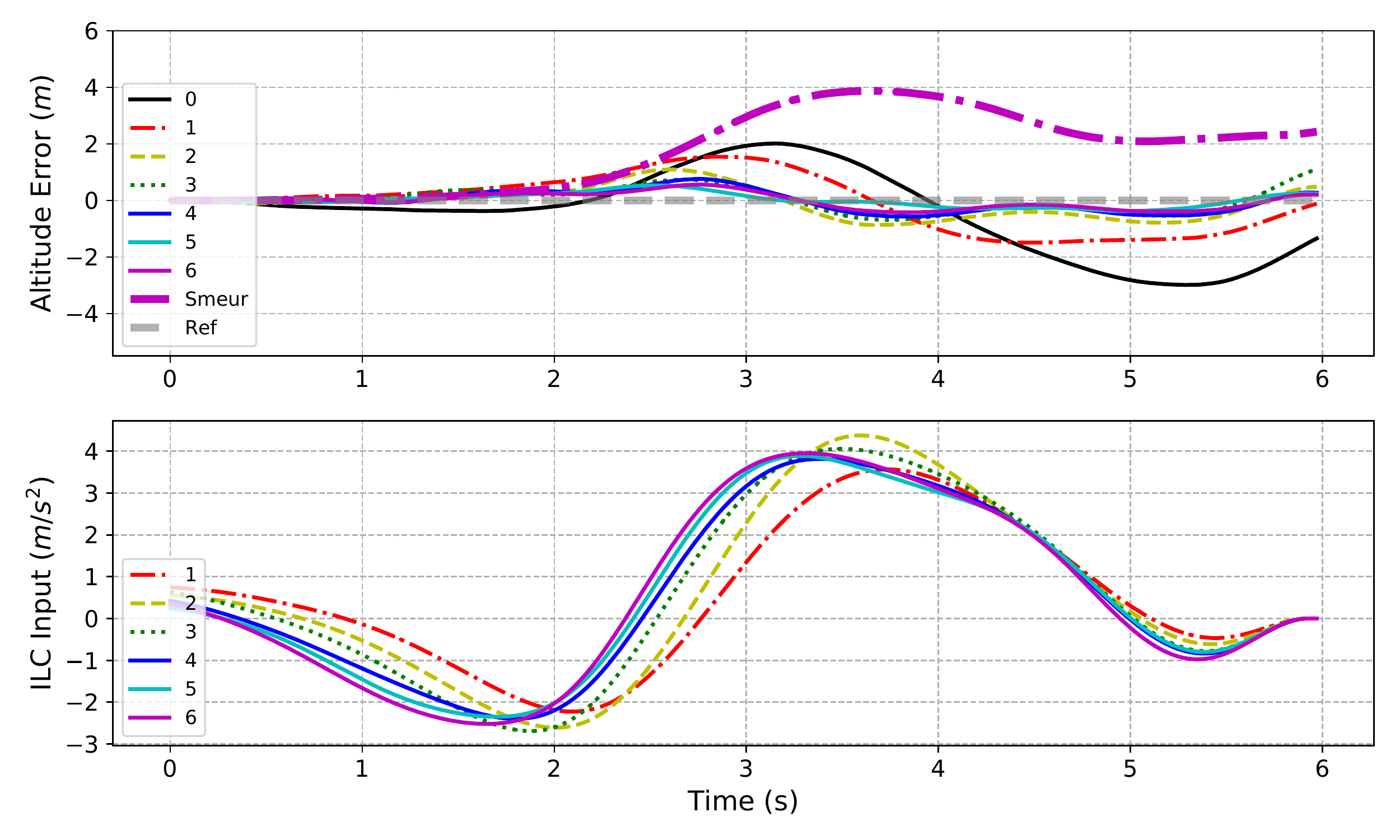}}
		\end{center}
		\vspace{-0.5cm}
		\caption{\label{fig:ILC-test-10} Altitude error and ILC input during each iteration when the head-up angle is $70\,^\circ$.}
		\vspace{-0.25cm}
	\end{figure}

\begin{figure}[h]
		\vspace{0.cm}
		\begin{center}
			{\includegraphics[width=1.0\columnwidth]{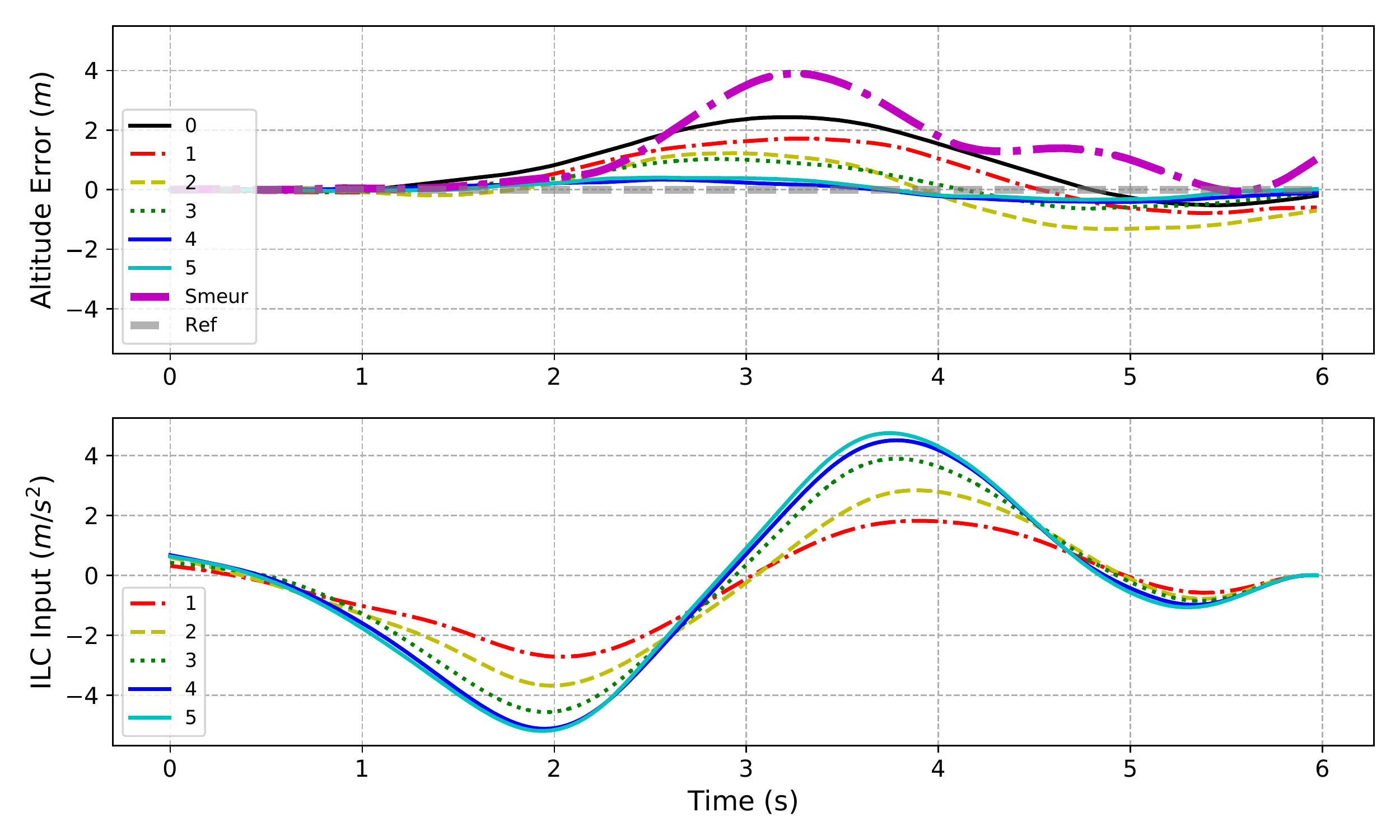}}
		\end{center}
		\vspace{-0.5cm}
		\caption{\label{fig:ILC-test-30} Altitude error and ILC input during each iteration when the head-up angle is $50\,^\circ$.}
		\vspace{-0.5cm}
	\end{figure}

	Fig.\ref{fig:ILC-test}, \ref{fig:ILC-test-10} and \ref{fig:ILC-test-30} show the altitude error and the ILC control actions for each iteration of the ILC algorithm. As can be seen, in all these three maneuvers, the ILC converges in four or five iterations. At convergence, the altitude error root mean square (r.m.s.) are as small as $10\,cm$, $23\,cm$ and $23\,cm$, respectively. These results show that our proposed control methods can effectively attenuate the altitude error in Cobra maneuver with different pitch angle profiles.
	
	\begin{figure}[h]
		\vspace{-0.0cm}
		\begin{center}
			{\includegraphics[width=1\columnwidth]{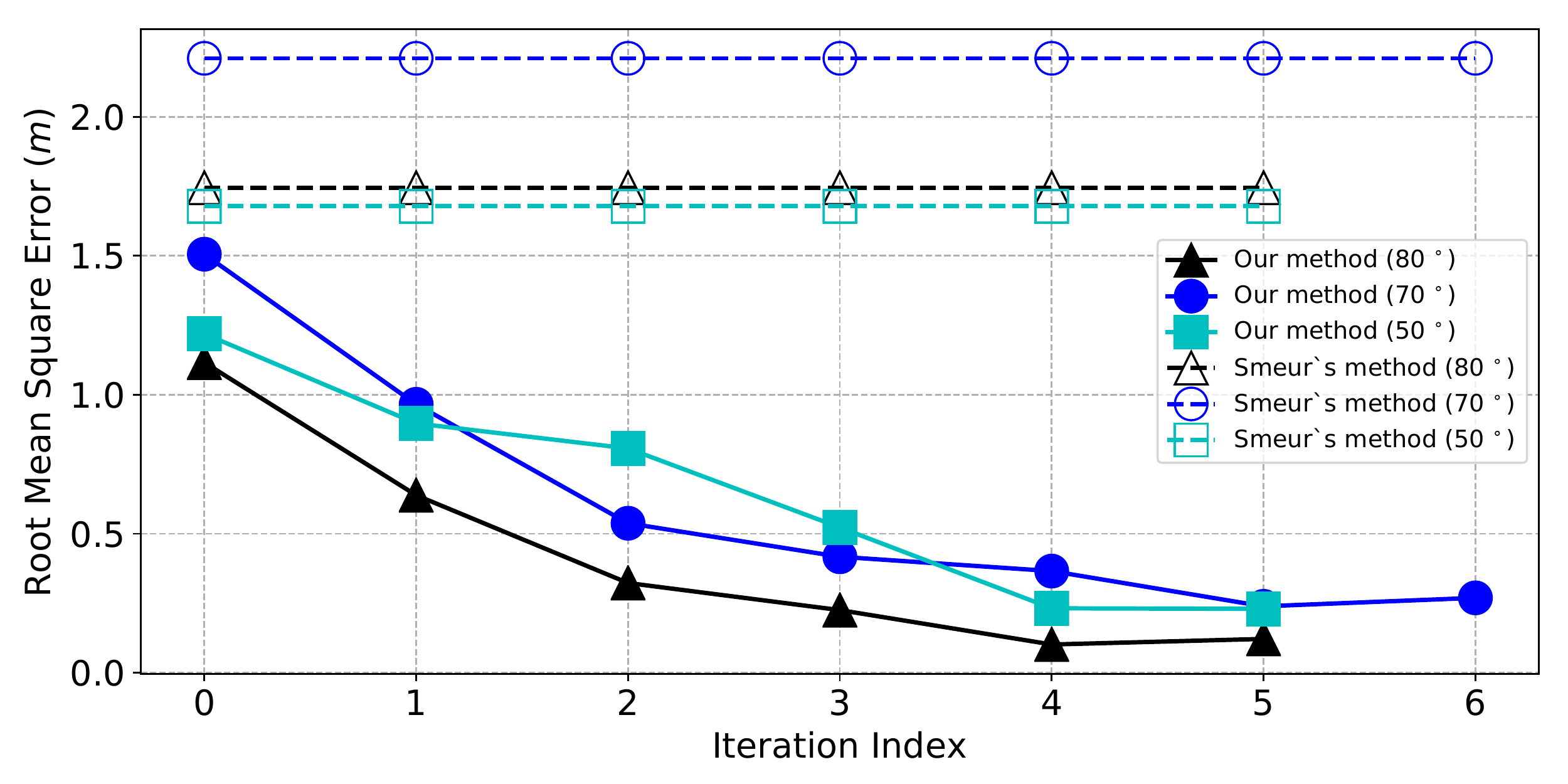}}
		\end{center}
		\vspace{-0.5cm}
		\caption{\label{fig:rmse} Root mean square error of altitude. The Smeur's method does not learn from iterations. So the same root mean square error is used for all iterations. }
		\vspace{-0.5cm}
	\end{figure}
	
	Furthermore, for all these three maneuvers, we conduct a comparison study with a recent method in\cite{smeur2019incremental}. Like our method, this method, referred to as {\it{Smeur's method}}, also uses acceleration measurements as feedback and does not need to identify the aircraft aerodynamic force coefficient, so a comparison against which would be much fair. The flight experiment results are shown in Fig. 15, where we can see that our control method based on the acceleration model surpasses the Smeur's method substantially in all these three Cobra maneuvers. Even without ILC, our baseline controller achieves an altitude change of $1.12\,m$ (versus $1.74\, m$, $35.63\,\%$ improvement), $1.51\, m$ (versus $2.21\, m$, $31.67\,\%$ improvement), $1.22\, m$ (versus $1.68 \,m$, $27.38\,\%$ improvement), respectively. The video of the final Pugachev's Cobra maneuver at convergence when the head-up angle is $80\,^\circ$ can be found at https://youtu.be/zUJEqgDy0RA.
	
	\section{Conclusion and Future work}
	To best exploit the maneuverability of tail-sitter UAVs, the Pugachev's Cobra maneuver is considered in this paper. For this maneuver, a complete control system has been proposed, which consists of two parts, a lateral controller and a feedback-feedforward altitude controller. Both are designed based on the UAV acceleration model. To compensate for the environmental disturbance and unmodeled dynamics, an iterative learning control algorithm is proposed. A big advantage of the acceleration based iterative learning control is that it does not require to identify the UAV aerodynamic parameters, eliminating the use of costly wind tunnel tests while achieving state-of-the-art control performance. The proposed control methods are verified by real-world outdoor flight experiments.
	
	A limitation of the proposed method is the singularity at $90\,^\circ$ pitch angle. Although our tail-sitter UAV normally flies with $8\,^\circ$ angle of attack (pitch angle $-82\,^\circ$) for the best aerodynamic efficiency and no singularity takes place in the Pugachev's Cobra maneuver. It prevents the tail-sitter UAV from further pitching down to achieve higher speed (at the cost of lower efficiency). Our future work will focus on extending this acceleration model-based control technique to the general control of tail-sitter UAV, which eliminates the need to model the UAV aerodynamics accurately.
	\bibliography{paper} 
\end{document}